\documentclass[aps,prb,twocolumn,superscriptaddress,showpacs]{revtex4}
\usepackage{graphics}
\usepackage{graphicx}
\usepackage[english]{babel}
\usepackage{color}
\usepackage{amsmath}
\usepackage{mathtools}

\begin{document}

\title{Anisotropic suppression of hyperuniformity of elastic systems in media with planar disorder}

\author{Joaqu\'{i}n Puig}%
\affiliation{Centro At\'{o}mico Bariloche and Instituto Balseiro,
CNEA, CONICET and Universidad Nacional de Cuyo, Avenida Bustillo
9500, 8400 San Carlos de Bariloche, Argentina}
\author{Federico El\'{i}as}
\affiliation{Centro At\'{o}mico Bariloche and Instituto Balseiro,
CNEA, CONICET and Universidad Nacional de Cuyo, Avenida Bustillo
9500, 8400 San Carlos de Bariloche, Argentina}
\author{Jazm\'{i}n Arag\'{o}n S\'{a}nchez}%
\affiliation{Centro At\'{o}mico Bariloche and Instituto Balseiro,
CNEA, CONICET and Universidad Nacional de Cuyo, Avenida Bustillo
9500, 8400 San Carlos de Bariloche, Argentina}
\author{Ra\'ul Cort\'es Maldonado}%
\affiliation{Centro At\'{o}mico Bariloche and Instituto Balseiro,
CNEA, CONICET and Universidad Nacional de Cuyo, Avenida Bustillo
9500, 8400 San Carlos de Bariloche, Argentina}
\author{Gonzalo Rumi}
\affiliation{Centro At\'{o}mico Bariloche and Instituto Balseiro,
CNEA, CONICET and Universidad Nacional de Cuyo, Avenida Bustillo
9500, 8400 San Carlos de Bariloche, Argentina}
\author{Gladys Nieva}
\affiliation{Centro At\'{o}mico Bariloche and Instituto Balseiro,
CNEA, CONICET and Universidad Nacional de Cuyo, Avenida Bustillo
9500, 8400 San Carlos de Bariloche, Argentina}
\author{Pablo Pedrazzini}
\affiliation{Centro At\'{o}mico Bariloche and Instituto Balseiro,
CNEA, CONICET and Universidad Nacional de Cuyo, Avenida Bustillo
9500, 8400 San Carlos de Bariloche, Argentina}
\author{Alejandro B. Kolton}
\affiliation{Centro At\'{o}mico Bariloche and Instituto Balseiro,
CNEA, CONICET and Universidad Nacional de Cuyo, Avenida Bustillo
9500, 8400 San Carlos de Bariloche, Argentina}
\author{Yanina Fasano$^{*}$}%
\affiliation{Centro At\'{o}mico Bariloche and Instituto Balseiro,
CNEA, CONICET and Universidad Nacional de Cuyo, Avenida Bustillo
9500, 8400 San Carlos de Bariloche, Argentina}

\date{\today}

\begin{abstract}

\section*{Abstract}

Disordered hyperuniform materials with vanishing long-wavelength
density fluctuations are attracting attention due to their unique
physical properties. In these systems, the large-scale density
fluctuations are strongly suppressed as in a perfect crystal, even
though the system can be disordered like a liquid. Yet,
hyperuniformity can be affected by the different types of quenched
disorder unavoidably present in the host medium where constituents
are nucleated. Here, we use vortex matter in superconductors as a
model elastic system to study how planar correlated disorder impacts
the otherwise hyperuniform structure nucleated in samples with weak
point disorder. Planes of defects suppress hyperuniformity in an
anisotropic fashion: While in the transverse direction to defects
the long-wavelength density fluctuations are non-vanishing, in the
longitudinal direction they are smaller and the system can
eventually recover hyperuniformity for sufficiently thick samples.
Our findings stress the need of considering the nature of disorder
and thickness-dependent dimensional crossovers in the search for
novel hyperuniform materials.

$^{*}$ Corresponding author: Yanina.Fasano@cab.cnea.gov.ar
\end{abstract}

 \maketitle

\section*{Introduction}

A great number of disordered physical and biological systems are
endowed with a universal hidden order characterized by a
macroscopically uniform density of
constituents.~\cite{Torquato2003,Torquato2018}  This  hidden order
is the structural property of hyperuniformity, characterized by an
anomalous suppression of large-scale density fluctuations in the
system. This property is naturally expected in a crystal, but it is
also observed in a wide variety of disordered systems such as
two-dimensional material
structures,~\cite{Man2013,Chen2018,Zheng2020,Salvalaglio2020,Chen2021,Chen2021b}
 jammed particles,~\cite{Zachary2011,Dreyfus2015}
bubbles in foam,~\cite{Chieco2021} vortex matter in type-II
superconductors,~\cite{Rumi2019,Llorens2020b} patterns of
photoreceptors in avian retinas,~\cite{Jiao2014} biological
tissues,~\cite{Zheng2020c} and even the distribution of the density
fluctuations in the early Universe.~\cite{Gabrielli2003}
Hyperuniform systems present a vanishing structure factor in the
infinite-wavelength  or small wavenumber $q$ limit, namely
$S(\mathbf{q}) \to 0$ as $\mathbf{q} \to
\mathbf{0}$.~\cite{Torquato2003,Torquato2018} This magnitude can be
directly measured via different diffraction techniques and provides
information on the fluctuations of the density of constituents of
the system at different wavenumbers. Since hyperuniformity is a
property defined in an asymptotic limit, strict hyperuniformity is
difficult to ascertain in experimental as well as computer-simulated
systems. Then, most works show that the systems are effectively
hyperuniform.~\cite{Klatt2019}

This exceptional but ubiquitous state of matter presents a
phenomenology that goes against the conventional wisdom on the
effect of disorder in the physical properties of  systems of
interacting objects.~\cite{Man2013,Chen2018,Torquato2018} For
instance,  disorder  typically lowers the electrical conductivity of
metallic materials. However, a recent work reports that disordered
hyperuniform systems present a  closing of  bandgaps resulting in an
enhanced conductivity.~\cite{Zheng2020} Also, periodic or
quasiperiodic order  was assumed as a prerequisite for a material to
present photonic bandgap properties. Strikingly, disordered
hyperuniform-engineered materials possess complete photonic bandgaps
blocking all directions and polarizations for short
wavelengths.~\cite{Florescu2009,Man2013,FroufePerez2016} In
addition, hyperuniform patterns can be very useful in practical
technological applications. For example, hyperuniform patterns of
 defects can pin with high efficiency the vortex structure
nucleated in superconductors, avoiding the undesirable dissipation
that can occur in superconducting
devices.~\cite{Lethien2017,Sadovskyy2019}

Theoretically, due to the fluctuation-compressibility theorem,
hyperuniformity may naturally emerge at thermal equilibrium in
incompressible systems with long-range repulsive interactions
between the constituents. \cite{Torquato2018} Nevertheless, a
hyperuniform point pattern within a higher dimensional system
presenting only short-range interactions or gradient terms at
equilibrium may also exist. Indeed, a three dimensional vortex
lattice model with short-range repulsions and local elasticity may
present  hyperuniform two-dimensional point patterns at every plane
perpendicular to the vortex lines.~\cite{Rumi2019} In general terms,
this road to hyperuniformity results from bulk-mediated  effective
long-range interactions between the points in the hyperuniform
pattern.

Vortex matter in superconductors is a model system to study the
occurrence of hyperuniformity in media with different types of
disorder. Vortices are elastic objects that nucleate in type-II
superconductors when applying a magnetic field. They are string-like
zones of the material that concentrate a quantized amount of
magnetic flux and interact repulsively between each other. The
competition between this repulsion and the pressure exerted by the
field results in vortices forming a structure with lattice spacing
$a\propto B^{-1/2}$. The vortex structure stabilizes in solid,
glassy and liquid phases, depending on temperature, applied field,
the particular material and the nature of disorder in the
samples.~\cite{Cubitt1993,Blatter1994,Pardo1998,Klein2001,Menghini2002,Pautrat2007,Petrovic2009,Suderow2014,MarzialiBermudez2015,Zehetmayer2015,ChandraGanguli2015,Toft-Petersen2018,AragonSanchez2019}
The nucleation of quasi-ordered and disordered hyperuniform vortex
structures has first been reported experimentally in samples of the
high-$T_{\rm c}$ Bi$_{2}$Sr$_{2}$CaCu$_{2}$O$_{8 + \delta}$  with
respectively weak point and strong columnar
disorder.~\cite{Rumi2019}  Later, disordered hyperuniform vortex
structures have been observed at high fields in several
superconductors  presenting weak and strong point
disorder.~\cite{Llorens2020b} Hyperuniform vortex structures are
theoretically expected for media with weak point disorder since in
this case the effective interaction between vortex tips at the
sample surface is long ranged. Theoretically, a suppression of
hyperuniformity is expected for media with columnar correlated
disorder.~\cite{Rumi2019} However, in the latter case an algebraic
decay of $S(\mathbf{q})$ in the $\mathbf{q} \to \mathbf{0}$ limit is
detected experimentally. This apparent discrepancy between theory
and experiment is quite likely due to the viscous freezing of the
system when field-cooling from the hyperuniform vortex liquid phase
towards the low-temperature glassy vortex phase.~\cite{Rumi2019} In
contrast, in the case of a type-II superconductor with planar
correlated defects,  strong fluctuations of the vortex density have
been proposed as the fingerprint of a disordered gel of
vortices.~\cite{Llorens2020} Thus, the nature of disorder in the
host medium plays a determinant role on the magnitude of density
fluctuations, and thus their effect on the nucleation of
hyperuniform materials deserves further investigation.

Here we address the question of whether planar correlated quenched
disorder, even if present in a reduced region of the sample, can
ultimately affect the hyperuniform hidden order. We use vortex
matter in two different high-$T_{\rm c}$ superconductors as model
systems. In order to be reliable, these studies on long-range vortex
density fluctuations require  high-resolution direct imaging of
individual vortices in extended fields-of-view with thousands of
vortices or more. Our experimental data with such a resolution and
extension are contrasted with numerical simulations of a system of
interacting elastic strings nucleated in media with planar disorder.
We show that planes of crystal defects running all the way through
the sample thickness produce a suppression of hyperuniformity in an
anisotropic fashion. Furthermore, we discuss how finite size effects
are relevant for these observations and how its removal in a
sufficiently-thick sample can produce a recovery of the hyperuniform
hidden order in the direction longitudinal to planar defects.

\section*{Results}

\subsection*{Density fluctuations on large length scales in media with planar defects}

A practical way to image vortex density fluctuations on large length
scales is to apply the magnetic decoration technique to take
snapshots of thousands of vortices at the sample surface. Magnetic
decoration consists in producing nanometer-size Fe clusters  that
are attracted towards the local field gradient entailed by  vortices
on the surface of the sample.~\cite{Fasano2008} We decorate vortex
positions at 4.2\,K after field-cooling; then the sample is warmed
up to room temperature and the Fe clusters are imaged by means of
scanning electron microscopy. We investigate
Bi$_2$Sr$_2$CaCu$_2$O$_{8+\delta}$ and YBa$_{2}$Cu$_{3}$O$_{7}$
high-$T_{\rm c}$ superconducting samples as model media with planar
correlated quenched disorder, namely domains with enhanced pinning.
In the first case we study samples with few and many planar defects
separating zones of the sample with slightly different orientations
of their c-axis;~\cite{Koblischka1995,Herbsommer2001} in the second
case we consider samples with twin-boundaries that act as planar
defects. For comparison, we study samples of both materials with
point disorder only, namely with no planar defects as revealed by
means of magnetic decoration. See Methods for further details on the
experimental techniques and sample characterization.

\begin{figure*}[hht]
    \includegraphics[width=2\columnwidth]{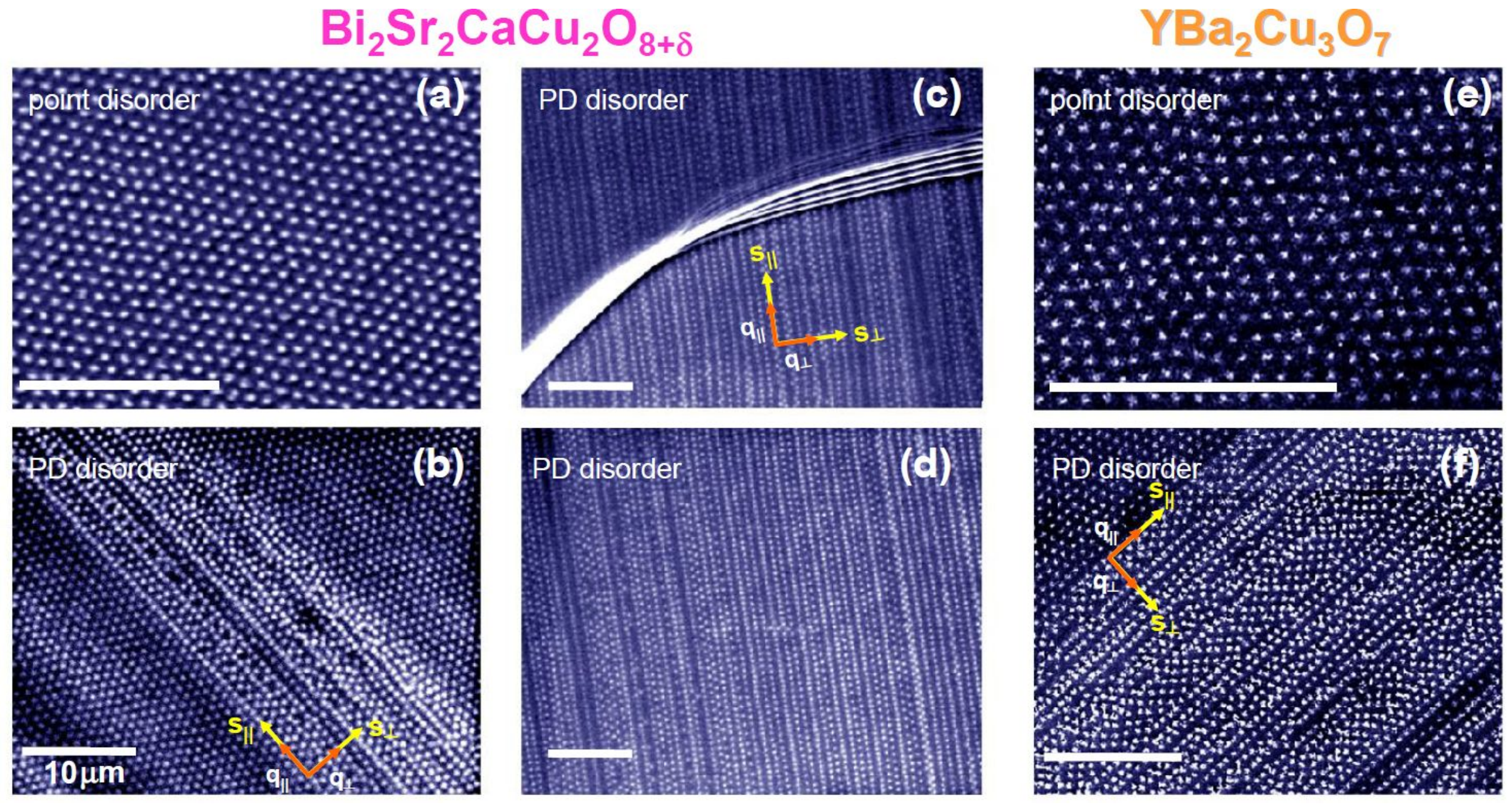}
    \caption{\textbf{Magnetic decoration imaging of vortices in samples with point and
    planar defects.} Vortices (white dots) nucleated at 30\,G in
     Bi$_2$Sr$_2$CaCu$_2$O$_{8+\delta}$ (a-d) and
     YBa$_{2}$Cu$_{3}$O$_{7}$ (e-f) samples with point (a,e)  and
     planar (b,c,d,f) disorder. Vortices nucleated on planar defects are seen as \textit{vortex rows} with a density in general larger than the average. Results in Bi$_2$Sr$_2$CaCu$_2$O$_{8+\delta}$ samples presenting few (b) and many (c,d) planar defects in millimeter-sized crystals. (c) Illustration on how the planar defects extend on the bulk of the sample: The vortex rows continue  in the same in-plane
    direction in the vicinity of a micron-height step generated by cleaving. The $s_{\parallel}$-$s_{\perp}$ coordinate system oriented along and perpendicular to the planar defects (PD), and the corresponding wave vectors $q_{\parallel}$
    and $q_{\perp}$, are shown in some images.  White scale bars
    correspond to $10\,\mu$m and labels indicating the dominant disorder are shown in
    each case.}
    \label{fig:Figure1}
\end{figure*}

Figure\,\ref{fig:Figure1}  shows images of the vortex structures
nucleated in some of the studied samples. In the case of point
disorder the structure is  hexagonal whereas the planar defects
induce the formation of vortex rows oriented along the direction of
defects $s_{\parallel}$, or correspondingly $q_{\parallel}$ in
reciprocal space. These vortex rows have generally a larger  density
than the average, in agreement with evidence from different imaging
and dynamic
techniques~\cite{Koblischka1995,Herbsommer2001,Maggio-Aprile1997,Fasano1999}
that indicate that planar defects in both materials act as strong
pinning centers for vortices. In Bi$_2$Sr$_2$CaCu$_2$O$_{8+\delta}$
samples with few planar defects, vortex rows are observed in a
micron-sized region, and the hexagonal structure is recovered
elsewhere, see Fig.\,\ref{fig:Figure1} (b). The sample with many
planar defects was specially chosen since most vortices are aligned
in vortex rows in the whole crystal, see Figs.\,\ref{fig:Figure1}
(c) and (d). In this sample, as also reported in samples with few
planar defects,~\cite{Herbsommer2001,Fasano1999} the alignment of
the rows is not altered by surface steps resulting from cleaving,
indicating the defects extend towards the bulk of the crystals. In
the case of YBa$_{2}$Cu$_{3}$O$_{7}$, the structure is hexagonal in
the untwinned sample and vortex rows are also observed in a heavily
twinned sample, see Figs.\,\ref{fig:Figure1} (e) and (f).

\begin{figure}[ttt]
    \includegraphics[width=\columnwidth]{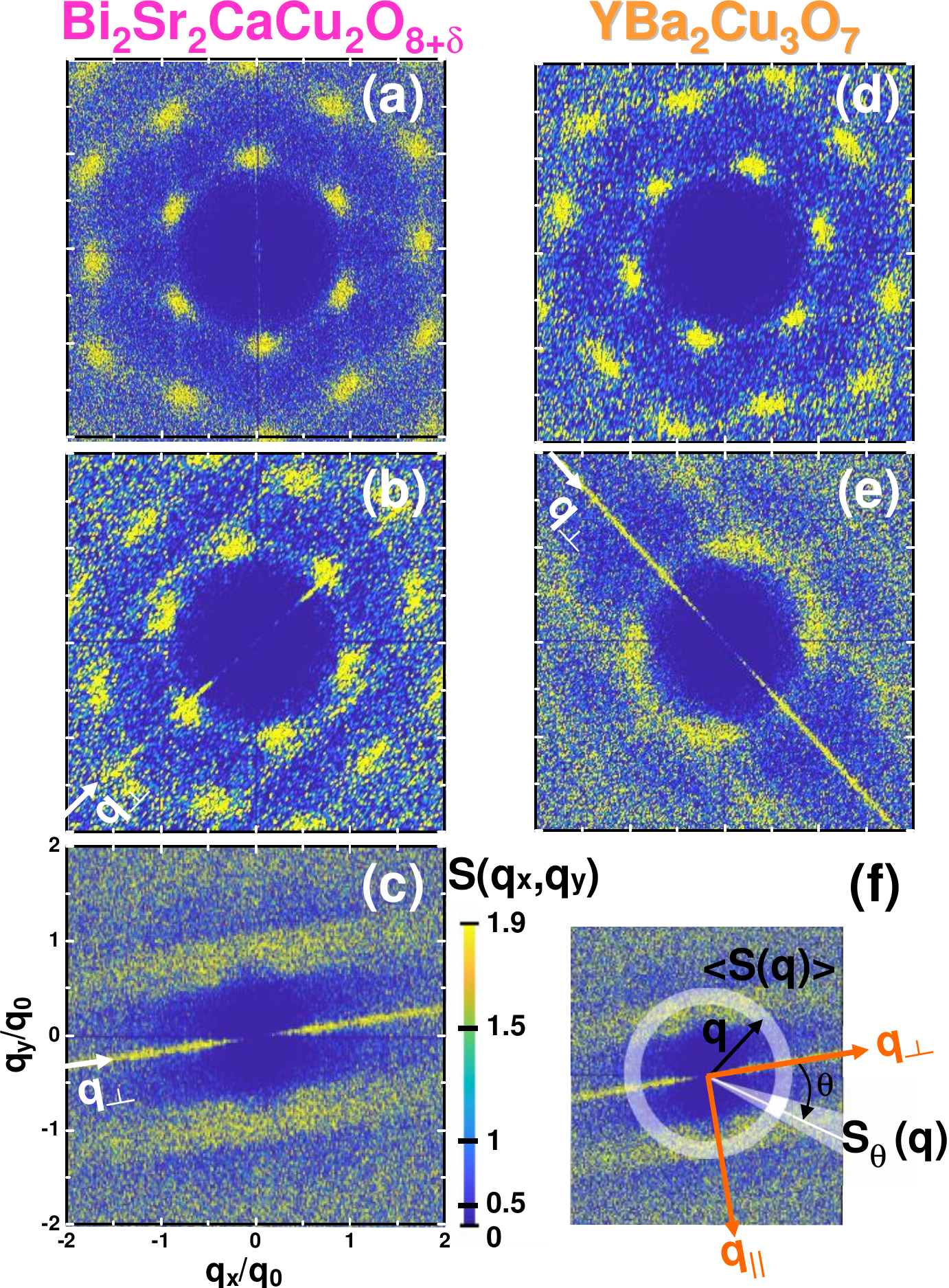}
    \caption{\textbf{Structure factor of vortices nucleated
        in type-II superconducting
        samples with point and planar disorder.}
Data in Bi$_2$Sr$_2$CaCu$_2$O$_{8+\delta}$ samples with (a) point
disorder, (b)  few and (c) many planar defects (9,200, 2,000  and
35,000 vortices). Data in YBa$_{2}$Cu$_{3}$O$_{7}$ samples with (d)
point disorder and (e)  many twin boundaries (2,300 and 4,000
vortices). The color-scale is cuadratic (see bar) and wavenumbers
are normalized by the Bragg wavenumber $q_{0}=2\pi/a$ with $a$ the
average distance between first-neighbor vortices. (f) Schematics of
the computation of the angularly-averaged $\langle S(q) \rangle$ and
angular $S_{\theta} (q)$ structure factors: Pixels considered to
obtain these magnitudes are highlighted as a donut and an angular
section, respectively. Shown are the wave-vector $q$, angle
$\theta$, and the reciprocal-space directions $q_{\parallel}$ and
$q_{\perp}$. } \label{fig:Figure2}
\end{figure}

\begin{figure*}[ttt]
    \includegraphics[width=2\columnwidth]{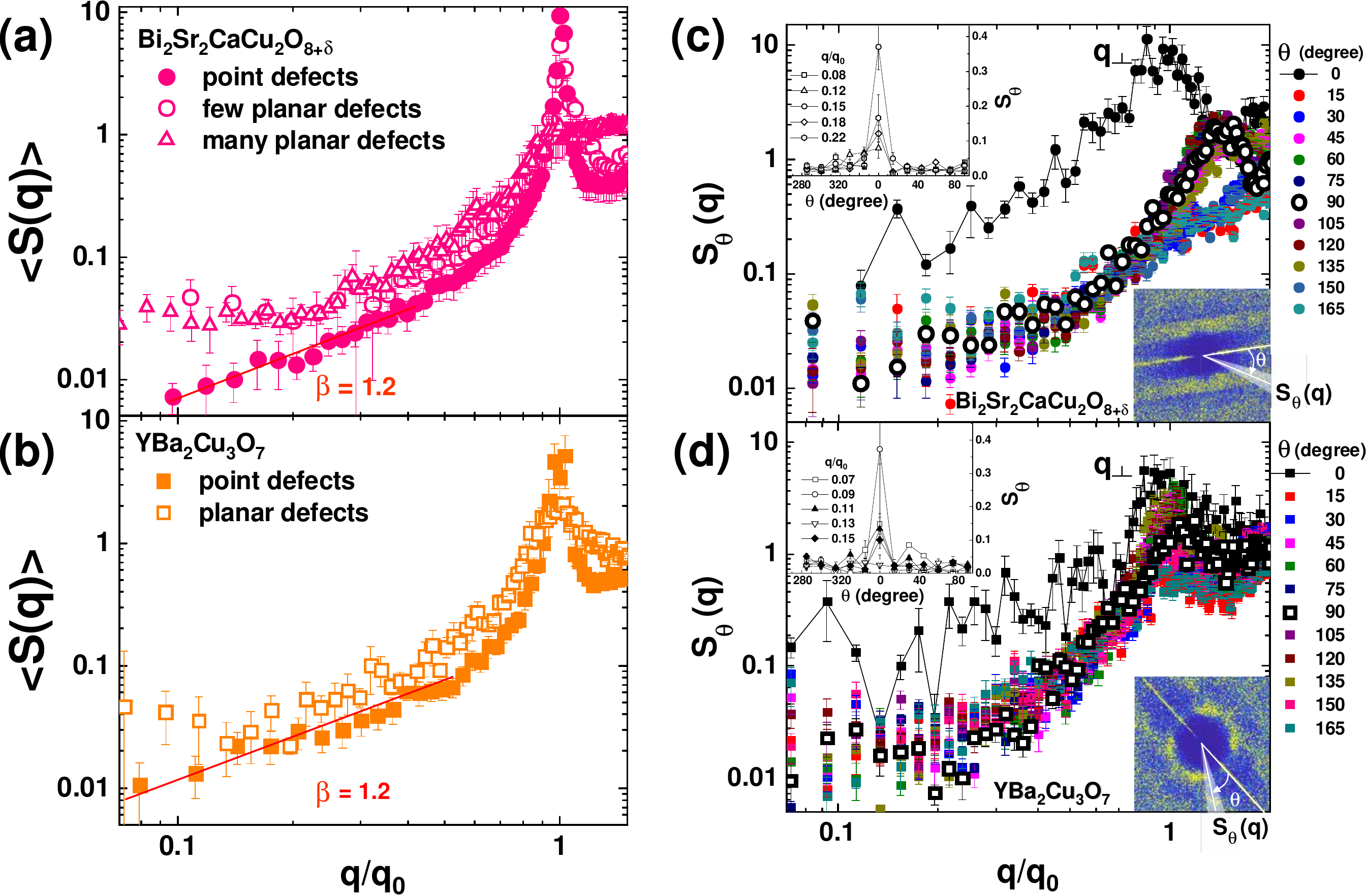}
    \caption{\textbf{Evidence supporting the suppression of hyperuniformity in
    media with planar defects.} Angularly-averaged structure factor of the vortex structures nucleated
        in (a) Bi$_2$Sr$_2$CaCu$_2$O$_{8+\delta}$ and (b) YBa$_{2}$Cu$_{3}$O$_{7}$ samples  with point (full symbols) and planar (open symbols) defects. The wavenumber is normalized
by the Bragg wavenumber $q_{0}=2\pi/a$. Full red lines are fits to
the data in samples with point disorder considering $\langle
S(q)\rangle \sim (q/q_{0})^{\beta}$ when $q/q_{0} \rightarrow 0$
yielding $\beta = 1.2 \pm 0.2$. Angular $S_{\theta} (q)$ structure
factor of vortex matter nucleated in (c)
Bi$_2$Sr$_2$CaCu$_2$O$_{8+\delta}$ and (d) YBa$_{2}$Cu$_{3}$O$_{7}$
samples with planar defects. Curves obtained averaging the
$S(q_{x},q_{y})$ data shown in the inserts considering $\pm \delta
\theta=7.5$\,degree around the azimuthal angles $\theta$ indicated
in the legends. Black points correspond to $S_{\theta=0} (q)$ in the
transversal direction $q_{\perp}$; white-with-black-edge points show
$S_{\theta=90} (q)$ in the longitudinal $q_{\parallel}$ direction;
color points correspond to intermediate directions. In the $q \to 0$
limit all curves tend to saturate, but the color and white points
form a bunch of data saturating at a smaller value than data for the
transversal direction. Left-top inserts: $S_{\theta}$ as a function
of $\theta$ for  fixed values of $q/q_{0}$ (vertical cuts of data
shown in the main panels). Error bars represent the standard
deviation of data when averaging at a given $q$.}
\label{fig:Figure3}
\end{figure*}

In order to characterize the vortex density fluctuations on extended
fields-of-view, we analyze the structure factor $S(\mathbf{q})\equiv
S(q_{\rm x},q_{\rm y})$ from snapshots of the vortex arrangements
taken at the surface of these samples. Figure\,\ref{fig:Figure2}
shows $S(\mathbf{q})= |\hat{\rho}(q_{\rm x},q_{\rm y},z=0)| ^{2}$,
with $\hat{\rho}$ the Fourier transform  of the local vortex density
modulation at the surface  of the studied samples with typical
thickness $t \sim 5\,-\,30$\,$\mu$m.~\cite{Fasano2003} A strong
anisotropy is evident for samples with planar defects:
Figs.\,\ref{fig:Figure2} (b), (c)  and (e) show lines of local
maxima in $S(\mathbf{q})$ extended along the ${q_{\perp}}$ direction
(angle $\theta=0$) corresponding to vortex density fluctuations
transversal to planar defects. At first sight, the intensity seems
to faint on going towards $\bf{q}\rightarrow \bf{0}$. In the case of
media with point disorder,  $S(\mathbf{q})$ decays algebraically
when $q \rightarrow 0$, as reported
previously,~\cite{Rumi2019,Llorens2020} and also shown in
Figs.\,\ref{fig:Figure3} (a) and (b).  We wonder whether in samples
with planar defects this fainting is produced by an algebraic decay
in the $\bf{q}\rightarrow \bf{0}$ limit as expected for disordered
hyperuniform systems.

Figure\,\ref{fig:Figure3} shows one of the main findings of this
paper: The suppression of effective hyperuniformity  induced by the
addition of correlated planar disorder to the host medium. In
samples with planar defects, the angularly-averaged structure factor
$\langle S(q) \rangle$  for $q \to 0$ is larger than that for
samples with point defects (for definition of this magnitude, see
Fig.\,\ref{fig:Figure2} (f) and Methods). More significantly,
$\langle S(q) \rangle$ tends to saturate in the low-$q$ limit.  This
phenomenology is observed for the two studied compounds and for
crystals presenting few or many planar defects. In contrast, in
samples of the same compounds but with point disorder, $\langle S(q)
\rangle \sim q^{\beta}$  with $\beta \approx 1.2$ when $q
\rightarrow 0$, a signature of effective disordered hyperuniformity.

\subsection*{Anisotropy in the density fluctuations on large length
scales in samples with planar defects}

Here we show that in samples with planar defects the suppression of
hyperuniformity is anisotropic, with vortex density fluctuations of
greater magnitude in the $q_{\perp}$ than the $q_{\parallel}$
direction. First, we show that the saturation of the structure
factor in the $q \rightarrow 0$ limit is anisotropic for vortices
nucleated in samples with planar defects. Figures\,\ref{fig:Figure3}
(c) and (d) show data of the angular structure factor
$S_{\theta}(q)$, for different reciprocal space directions. Curves
with black (white with black edge) points correspond to $S_{\theta}
(q)$ data in the transverse direction $q_{\perp}$ (parallel
direction $q_{\parallel}$), whereas color points are data for
intermediate angles. Irrespective of the direction, at low $q$ all
curves tend to saturate, but while the color and white points form a
pack of data around $\sim 0.02-0.03$, the black curves corresponding
to the transverse modes $S_{\theta = 0} (q)$ stand out and saturate
at a value between 2 and 10 times larger. This is better depicted in
the inserts to Fig.\,\ref{fig:Figure3} (c) and (d). In addition, the
peaks in $S_{\theta} (q)$ are detected at smaller $q$ for $\theta=0$
than for $90$\,degree, indicating that the average vortex spacing is
smaller in the longitudinal than in the transverse direction to
planar defects.

\begin{figure*}[ttt]
    \includegraphics[width=2\columnwidth]{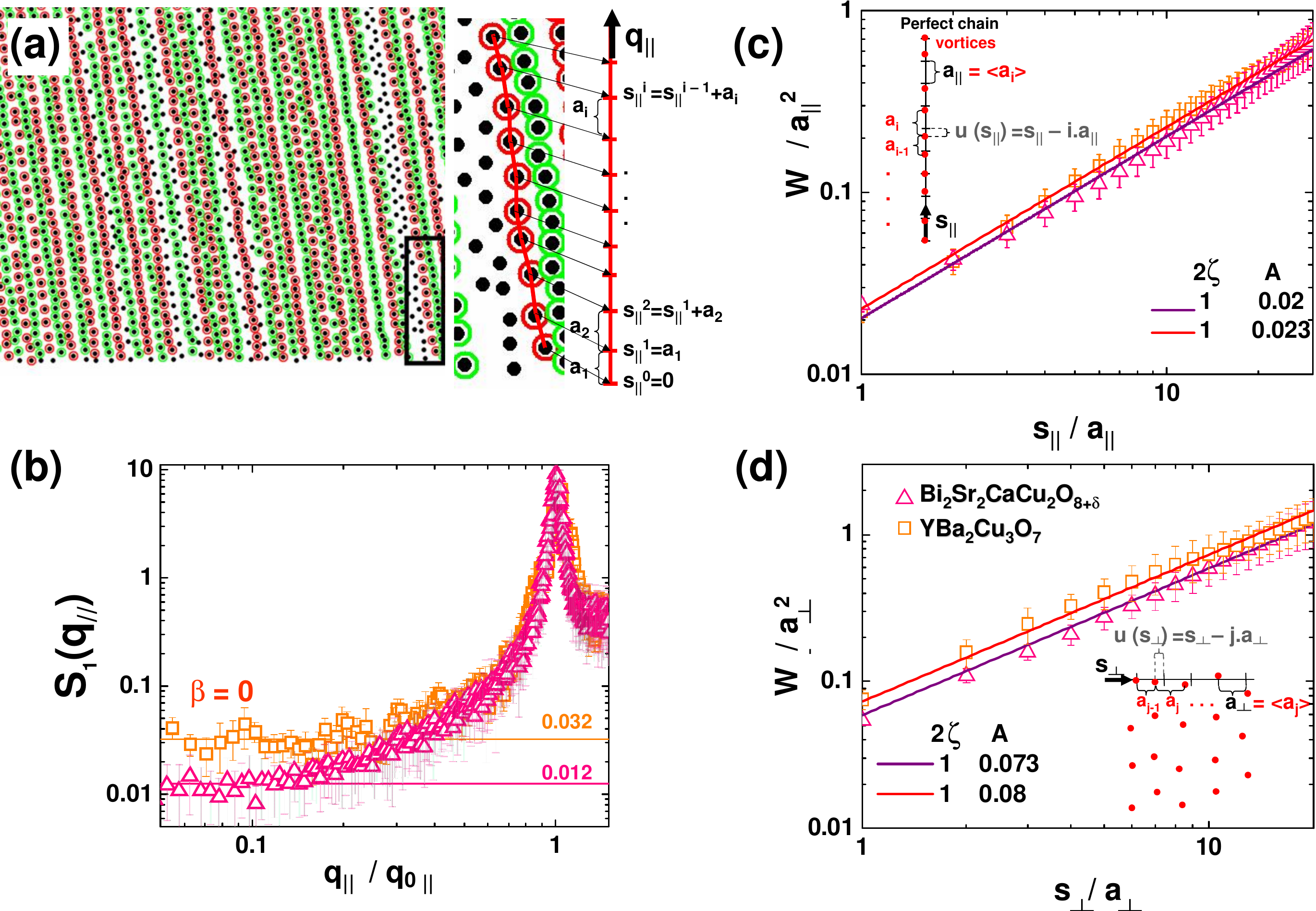}
    \caption{\textbf{Anisotropy in the large-scale  vortex density fluctuations
    in samples with planar defects.} (a) Left: Detail of the digitalized vortex positions (black dots) in
a region  of the Bi$_2$Sr$_2$CaCu$_2$O$_{8+\delta}$ sample with many
planar defects. Vortices highlighted in red or green belong to the
same vortex row; non-highlighted vortices are presumably
interstitial to the vortex rows. Right: Zoom-in of the structure in
the black-framed region. In a given row, vortices are labeled and
$a_{\rm i}$ is the distance between vortices $i$ and $i+1$. Then the
vortex row is mapped in a straight line where adjacent vortices are
spaced in $a_{\rm i}$ and the coordinate of vortex $i+1$ is
$s_{\parallel}^{\rm i}$. (b) After performing this mapping of vortex
    positions in a straight line, the average one-dimensional structure
    factor along the direction of defects, $S_{\rm
    1}(q_{\parallel})$ is computed (see text for details) averaging data coming from $\sim 50$ (12) rows containing 9,000 (1,300) vortices in Bi$_2$Sr$_2$CaCu$_2$O$_{8+\delta}$ ( YBa$_{2}$Cu$_{3}$O$_{7}$ ).   Fits considering an algebraic decay of $S_{\rm
    1}(q_{\parallel})$ for $q_{\parallel} \rightarrow 0$ yield $\beta=0$ within the error, for both materials. (c) Displacement correlator
    $W/a_{\parallel}^{2}$ calculated along the $s_{\parallel}$ direction of vortex rows (statistics for 1,500 vortices in Bi$_2$Sr$_2$CaCu$_2$O$_{8+\delta}$ and  1,300 in YBa$_{2}$Cu$_{3}$O$_{7}$).  Insert: Schematic representation of the magnitudes considered for the calculation of $W$: $u (s_{\parallel})= s_{\parallel} - i \cdot a_{\parallel}$ is the displacement of the $i$-th vortex  from the site of a perfect vortex chain with lattice spacing $a_{\parallel}= \langle a_{i} \rangle$, the average in a row. (d) Displacement correlator
    $W/a_{\perp}^{2}$ calculated along the transversal direction $s_{\perp}$  (statistics for 1,300 vortices in Bi$_2$Sr$_2$CaCu$_2$O$_{8+\delta}$ and
 1,200 in YBa$_{2}$Cu$_{3}$O$_{7}$).
    Insert: Schematic representation on the magnitudes considered for
     the computation of $W$ with $a_{\perp}= \langle a_{j} \rangle$, the average
     in a line. Fits of the displacement correlators with algebraic functions
     $A\cdot x^{2\zeta}$ (full lines) yield the roughening exponents $2\zeta$ and
     factors $A$ indicated in the legends. Error bars represent the standard deviation of data when averaging at a given $q$.}
\label{fig:Figure4}
\end{figure*}

Second, in order to better characterize this anisotropy, we consider
the one-dimensional structure factor of individual vortex rows that
is sensitive to vortex density fluctuations along the direction of
planar defects. In order to compute this magnitude, the experimental
positions of vortices in  individual rows are mapped in a straight
line  such that adjacent vortices are spaced a distance $a_{\rm i}$
and the coordinate of vortex $i+1$ is  $s_{\parallel}^{\rm i}$, see
the schematic representation  of Fig.\,\ref{fig:Figure4} (a).  The
one-dimensional structure factor of a given row is then computed as
$|\hat{\rho}_{1}(q_{\parallel})|^{2}$, with $\hat{\rho_{1}}$ the
Fourier transform  of the vortex density modulation along the line.
Then, for each vortex row we calculate the average lattice spacing
in a row, $a_{\parallel}\equiv\langle a_{i} \rangle$, and the
wavenumber $q_{\parallel}$ is normalized by $q_{0\,\parallel}\equiv
2\pi/a_{\parallel}$. Finally, we average the one-dimensional
structure factor over many rows to obtain the $S_{\rm
1}(q_{\parallel})$ data shown in Fig.\,\ref{fig:Figure4} (b). In the
$q \rightarrow 0$ limit the tendency  to saturation is evident: A
fit of $S_{\rm 1}(q_{\parallel}) \sim q_{\parallel}^{\beta}$ yields
$\beta \approx 0$ within the error for the two studied compounds.

Third, we analyze the distance-evolution of the one-dimensional
displacement correlator computed along and perpendicular to the
direction of defects, $W(s_{\parallel})$ and $W(s_{\perp})$. These
magnitudes quantify the average over disorder of the displacement of
vortices in the $s_{\parallel}$ or $s_{\perp}$ directions with
respect to the sites of perfect chains oriented in the same
directions. The lattice spacing of the perfect chains, either
$a_{\parallel}$ or $a_{\perp}$, are obtained by averaging the
separation of vortices along the considered direction in a given
row/lane of vortices, see the schematics in the inserts to
Figs.\,\ref{fig:Figure4} (c) and (d).  In practice,  we compute
$W(s) = \langle [u(s) - u(0)]^{2} \rangle - \langle [u(s) - u(0)]
\rangle^{2}$, where $\langle ... \rangle$ is the average when
considering every vortex as the origin. $u$ is the displacement of
the $i$-th vortex located at $s_{\parallel}$($s_{\perp}$) from the
$i$-th site of a perfect chain of vortices with lattice spacing
$a_{\parallel}$ ($a_{\perp}$)  averaged in a given row(lane).

For disordered elastic systems, the displacement correlator
typically grows with distance with a roughness exponent $2\zeta$
given by the universality-class of the system along the considered
direction.~\cite{Barabasi1995} Figure\,\ref{fig:Figure4} (c) shows
the dependence of $W$  with $s_{\parallel}$ averaged over tens of
vortex rows and panel (d) shows the evolution with $s_{\perp}$
averaged along tens of vortex lanes. For the two studied compounds,
the data are reasonably well fitted with an algebraic growth with
exponent $2\zeta \approx 1$ in both directions, at odds with
expectations for a hyperuniform system. The  fits yield a
multiplicative factor $A$ roughly three and a half times larger for
the $s_{\perp}$ than for the $s_{\parallel}$ direction.  This is
another proof that from experimental evidence vortex density
fluctuations are anisotropic in media with planar defects.

\subsection*{Simulations of a structure of interacting elastic vortex lines in media with planar defects}

\begin{figure*}[ttt]
    \centering
    \includegraphics[width=2\columnwidth]{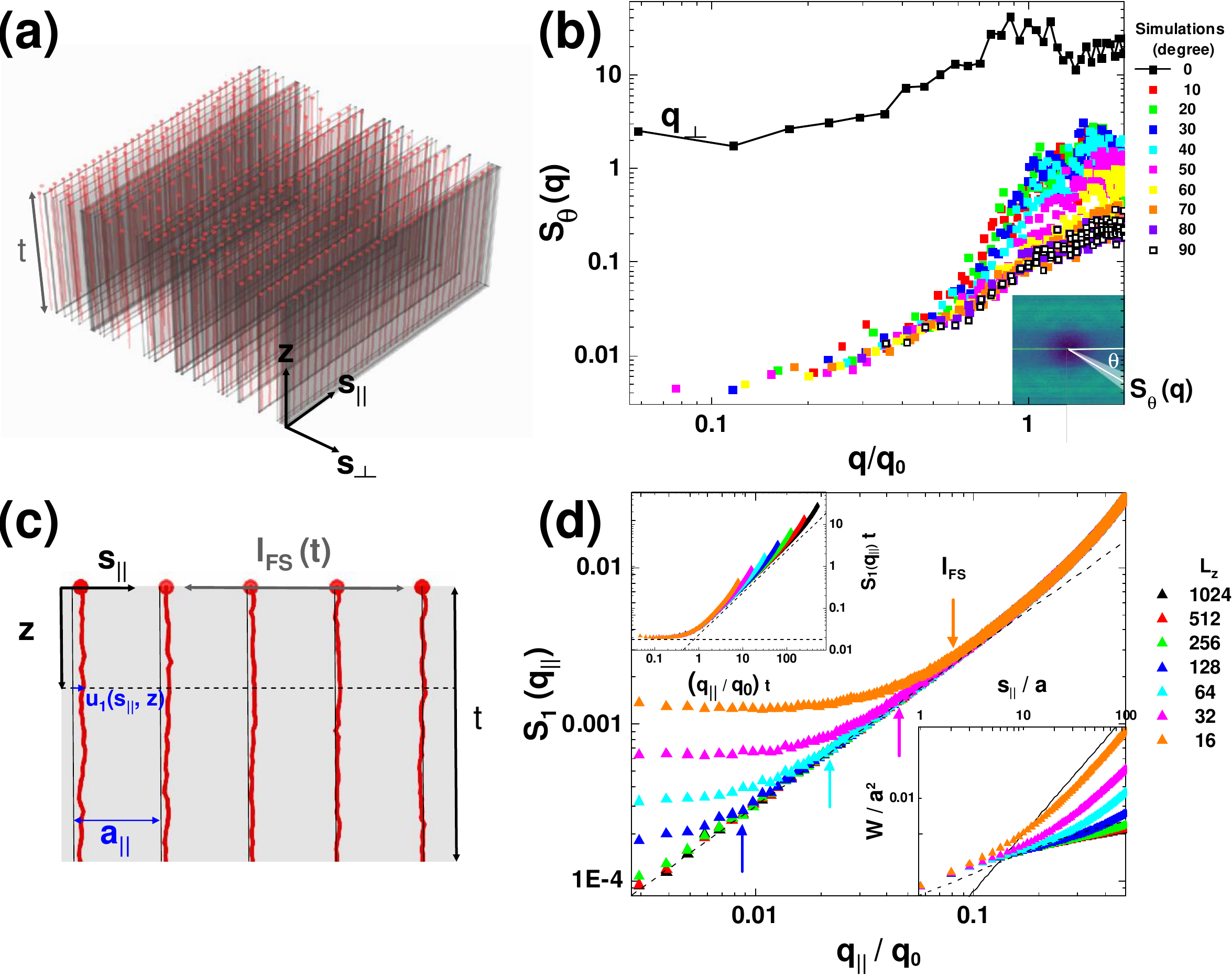}
    \caption{\textbf{Simulations of a structure of interacting elastic vortex lines
    in samples with planar defects.} (a) Snapshot of a quenched configuration obtained
    from three dimensional Langevin dynamics simulations of interacting vortices
    (red lines) in presence of parallel randomly-located  PD (gray planes). Circles
    highlight vortex tips at the sample surface. The coordinate system and the
    thickness are indicated. (b) Angular structure factor $S_{\theta}(q)$ of the
    vortex positions at the sample surface for different angles $\theta$ indicated
    in the legend. The insert shows the $S(q_{x},q_{y})$ data considered to calculate
    the curves in the main panel. (c) Schematics of the planar vortex row model
    indicating the displacement field $u_1(s_\parallel,z)$, the average spacing
    $a_{\parallel}$ for a perfect row with the same vortex density, and the
    characteristic thickness-dependent crossover length $l_{\rm FS}$. (d) Results
    for the configurations of the planar vortex row at the surface of the sample
    ($z=0$) for different sample thicknesses proportional to the number of layers
    in the simulation $L_{\rm z}$. Main panel:  $S_{1}(q_{\parallel})$ structure
    factor for different sample thicknesses. The dashed black line is a linear
    function. Arrows indicate the crossover behavior at
    $q_{\parallel}/q_{0} \sim 1/ l_{\rm FS}$. Bottom-right insert: Displacement
    correlator as a function of the distance along the row  $s_{\parallel}$. The
    full line
    is a linear function that reasonably describes the data in the large wavelength
    limit of thin samples whereas the dashed line is a logarithmic growth that
    follows the data at short wavelengths for thick enough samples. Top-left insert:
    Structure factor data $S_{1}(q_\parallel)t$ vs. $q_\parallel t$ showing a scaling
    collapse and the two regimes separated by the crossover wavevector,
    $2\pi/l_{\rm FS}\propto 2\pi/t$.}
    \label{fig:sims}
\end{figure*}

Here we gain insight on the origin of the anisotropic vortex density
fluctuations in media with planar defects by means of Langevin
dynamics simulations of vortex lines in three dimensions with an
applied field in the $z$-direction. We consider a media with
randomly-spaced parallel planar defects oriented with their normal
pointing along the $s_{\perp}$-axis in an orthogonal coordinate
system $(s_{\perp},s_{\parallel},z)$. We model $N_{\rm v}$ vortices
as elastic lines discretized in the $z$-direction, such that ${\bf
r}_{i}(z) \equiv (s_{\perp,i}(z),s_{\parallel,i}(z))$ describe the
two-dimensional coordinate of vortex $i$ at the layer
$z=1,...,L_{\rm z}$ with $L_{\rm z}$ the total number of layers.
Periodic boundary conditions are taken in all directions in a system
of size $L_{\perp} \times L_{\parallel} \times L_{\rm z}$. The total
energy per unit length of the structure of elastic lines is
$E[\{{\bf r}_{i}(z)\}]=E_{\rm v} + E_{\rm vv}+E_{\rm PD}$. Each line
has an elastic tension energy given by Hook couplings of strength
$k$
\begin{equation}
E_{\rm v}[\{{\bf r}_{i}(z)\}]=\sum_{i=1}^{N_{\rm v}}
\sum_{z=1}^{L_{\rm z}} \frac{k}{2} |{\bf r}_{i}(z+1)-{\bf
r}_{i}(z)|^2, \label{eq:tensionv}
\end{equation}
with $k = \epsilon_0/\lambda_{\rm ab}^2$
 a local harmonic approximation for the single vortex elastic tension,
$\epsilon_0 \equiv \phi_0^2/(8\pi^2 \lambda_{\rm ab}^2)$ the
interaction energy-scale per unit length, and $\lambda_{\rm ab}$ the
in-plane penetration length. The repulsive interaction energy
between three-dimensional vortex-lines derived from the London model
\cite{Blatter1994} is
\begin{equation}
E_{\rm vv}[\{{\bf r}_{i}(z)\}]=\sum_{i \neq j}^{N_{\rm v}}
\sum_{z=1}^{L_{\rm z}} {\epsilon_0} K_0 \left( \frac{|{\bf
r}_{j}(z)-{\bf r}_{i}(z)|}{\lambda_{\rm ab}} \right),
\label{eq:intvv}
\end{equation}
with $K_n(x)$ the nth-order modified Bessel function of the second
kind. The pinning energy due to $N_{\rm d}$ defects is modeled as
Gaussian-well channels
\begin{equation}
    E_{\rm PD}[\{{\bf r}_{i}(z)\}] = -A_{\rm pin} \sum_{n=1}^{N_{\rm d}} \sum_{i=1}^{N_{\rm v}} \sum_{z=1}^{L_{\rm z}} e^{-\frac{(s_{\perp,i}(z)-X_n)^2}{2\xi_{\rm ab}^2}}
    \label{eq:intpd}
\end{equation}
where $\xi_{\rm ab}$ is the in-plane coherence length, $A_{\rm pin}$
the pinning strength of the planar defects, and $X_n$ the random
positions of the planar defects uniformly sampled along $L_{\perp}$.
Finite-temperature Langevin dynamics simulations of the system are
performed to obtain equilibrated low temperature configurations (see
Methods). A snapshot of a configuration is shown in
Fig.\ref{fig:sims} (a).

Figures\,\ref{fig:sims} (a) and (b) show the main results of the
simulations that are in accordance with experimental observations:
i) The $S_{\theta}(q)$ is anisotropic and displays similar density
fluctuations for all $\theta \neq 0$ directions; ii) fluctuations in
the $q_{\perp}$ ($\theta=0$) transverse direction are orders of
magnitude larger, particularly at low $q$. The peak in the
transverse direction is detected at a smaller wavenumber than in
other directions, and $a_{\perp} > a_{\parallel}$ is also found in
the simulations.  Thus, this model of a structure of interacting
elastic vortex lines nucleated in planar defects that are in
controlled positions allow us to ascertain that the pinning
generated by defects is strong enough as to increase the vortex
density inside defects above the average. Furthermore, this model
reveals that the anisotropic suppression of hyperuniformity has
origin in the interactions allowing important vortex density
fluctuations at large wavelengths for vortices caged in defects but
also allowing for a rarefied distribution of vortices in between
defects. As discussed in detail in the next section, by performing
simulations and analytical calculations of a simplified version of
the model of Eqs.(\ref{eq:tensionv}),(\ref{eq:intvv}) and
(\ref{eq:intpd}) we can go further in the comparison between
experiments and theory, and show that the number of layers $L_{\rm
z}$ (proportional to the sample thickness), plays a very relevant
role in assessing hyperuniformity.

\section*{Discussion} \label{sec:disscusions}

The suppression of disordered hyperuniformity in media with planar
defects can be discussed in a broader context than that of its
implications for the synthesis of hyperuniform materials.  This
issue is connected to the related problem of the  structural phases
stabilized in media with different types of disorder. In the case of
planar defects oriented in the direction of the magnetic flux as we
study here, the stabilization of  a robust planar-glass phase is
expected.~\cite{emig2006,petkovic2009} In this phase, the positional
correlation function is expected to decay
exponentially,~\cite{emig2006,petkovic2009} implying both, a
displacement correlator function $W \sim s_{\perp}$, and a structure
factor behaving as $S_{\theta=0}(q\to 0)=\text{const}\neq 0$. These
theoretical implications are consistent with our experimental and
theoretical findings on the suppression of hyperuniformity in the
direction transverse to planar defects. Nevertheless, these
theoretical works do not study the vortex density fluctuations in
the direction longitudinal to planar defects nor the  experimentally
relevant size-effects. The saturation at a finite value in the
longitudinal direction $S_{\theta=90}(q\to 0)$, appreciably smaller
than $S_{\theta=0}(q\to 0)$, is a subtle issue. Indeed, we argue
below that at low densities, the vortex structure confined in a
planar defect can be disordered hyperuniform provided the sample is
thick enough and the confinement is strong.

In order to sustain these claims, we start by highlighting some
relevant findings. First, both in experiments and simulations, the
vortex structure in samples with planar defects presents well
defined vortex rows. Simulations also show that at low temperatures
most of vortex rows are parallel to planar defects. Second, the
average vortex spacing along a row is appreciably smaller than in
between rows, for instance $a_{\parallel} \sim 0.7 a_{\perp}$ in the
experiments. This indicates that intra-row vortex-vortex
interactions are stronger than inter-row ones, motivating a
single-row-based mean-field-like phenomenological approach.

Then, to further sustain our claims based in an analytical insight
of the problem, we now consider a simplified model that captures the
essential physical ingredients of the problem:  We neglect the
interaction between vortex rows as well as transverse vortex
fluctuations, and  model the system as a non-interacting collection
of single vortex rows with strongly localized vortices inside a
planar defect. We also neglect for the moment the effect of quenched
point disorder since it is expected to be weaker than the planar
defect pinning. The thermally-equilibrated configuration of the
elastic system can then be obtained analytically in the elastic
approximation by using the displacement field
$u_1(s_{\parallel},z)$. This field describes the mismatch of the
planar vortex row with respect to a perfectly periodic chain of
straight vortices aligned  in the $s_{\parallel}$-direction, see
Fig.\,\ref{fig:sims}(c). Within this simple model, as detailed in
Methods, the large-wavelength density fluctuations at a single layer
$z$ give a structure factor

\begin{equation}
    {S_1}(q_{\parallel})  \sim q_{\parallel}^2 \langle |\hat{u}_1(q_{\parallel},z)|^2 \rangle \approx
    \left\{
    \begin{aligned}
    \frac{n_0 k_{\rm B} T}{\sqrt{c_{11}c_{44}}}q_{\parallel} \;\;\; q_{\parallel} > \frac{2\pi}{l_{\rm FS}}\\
    \frac{n_0 k_{\rm B} T}{c_{11} t} \;\;\; q_{\parallel} < \frac{2\pi}{l_{\rm FS}} \\
    \end{aligned}
    \right.
    \label{eq:1lane}
\end{equation}
where $\hat{u}_1(q_{\parallel},z)$ is the Fourier transform of the
displacement field, $c_{11}$ and $c_{44}$ are the compression and
tilt elastic modulii of the planar vortex system, $t$ is the sample
thickness, and $l_{\rm FS}$ is a relevant crossover length. Assuming
translation symmetry along $z$,  Eq.(\ref{eq:1lane}) implies that in
real space the displacement correlator in the longitudinal direction
to planar defects scales as
\begin{equation}
    W(s_{\parallel})
    \sim
    \left\{
    \begin{aligned}
     \log(s_{\parallel})\;\;\; s_{\parallel}<l_{\rm FS}\\
     s_{\parallel}\;\;\; s_{\parallel}>l_{\rm FS},
    \end{aligned}
    \right.
    \label{eq:1lanew}
\end{equation}
see Methods for further details. The result of a
$\log(s_{\parallel})$ roughness is not a surprise since the system
is essentially a thermally-fluctuating two-dimensional elastic
lattice. The novelty here is that to explain the experimental
observations it is necessary to consider the  finite-thickness
induced crossover distance
\begin{equation}
l_{\rm FS} \approx \frac{t}{2\pi} \sqrt{\frac{c_{11}}{c_{44}}}\;\;.
\label{eq:yfs}
\end{equation}

This crossover behaviour in  $S_1(q_{\parallel})$ and
$W(s_{\parallel})$ is also confirmed in numerical simulations of a
single vortex row confined in a planar defect for different
thicknesses, see Fig.\ref{fig:sims}\,(d). Furthermore, the top-left
insert of this figure shows that all $S_1(q_{\parallel})t$ vs.
$q_{\parallel} t$ curves collapse into a master-curve, confirming
quantitatively that $l_{\rm FS}\propto t$, Eq.(\ref{eq:yfs}).
According to Eq. (\ref{eq:1lane}), in the  thermodynamic limit
$l_{\rm FS}\to \infty$, and the vortex row at a constant-$z$ cross
section is class II hyperuniform since ${S_1}(q_{\parallel})\sim
q_{\parallel}^{\alpha}$ with $\alpha=1$.~\cite{Torquato2018} This
hyperuniformity class contrasts with more ordered class I
hyperuniform systems where $\alpha>1$.~\cite{Torquato2018} However,
in systems with finite thickness, Eq.(\ref{eq:1lane}) predicts a
crossover towards a non-hyperuniform behaviour for
$s_{\parallel}>l_{\rm FS}$. This corresponds to a dimensional
crossover from a two-dimensional to an effective one-dimensional
elastic system equivalent to an elastic chain composed by rigid
vortices. Interestingly, this phenomenology is closely related to
the crossover predicted for Luttinger liquids at a characteristic
thermal length.~\cite{ThierryBook}

A caveat in our model might be that we ignore that real samples have
weak point disorder. However, as shown in Methods, if this disorder
is considered, the  main results of Eqs.(\ref{eq:1lane}),
(\ref{eq:1lanew}) and (\ref{eq:yfs}) remain qualitatively valid.
Namely, for $s_{\parallel} > l_{\rm FS}$ $S_1(q_{\parallel} \to 0) =
\text{const}$ and $W(s_{\parallel}\to \infty) \sim s_{\parallel}$
while for $s_{\parallel} < l_{\rm FS}$, the structure is disordered
hyperuniform but class III ($\alpha <1$) instead of class II
($\alpha =1$).~\cite{Torquato2018}

Finally, we argue that a finite size effect is a plausible
explanation for the suppression of hyperuniformity observed in
experiments and simulations. Considering the vortex-vortex
interaction potential per unit length $U(x)\approx \epsilon_0
K_0(x/\lambda)$, with $\epsilon_{\rm l} \sim \epsilon_0$ the single
vortex elastic tension, the elastic constants of the planar vortex
system can be estimated as $c_{11} \approx a U''(a)$ and $c_{44}
\approx \epsilon_{\rm l}/a$.~\cite{nattermann2000} Thus, using these
approximations in Eq.\ref{eq:yfs}, the crossover length can be
estimated by considering only the sample thickness since we get
$l_{\rm FS}\approx (t/2\pi)(a/\lambda)
\sqrt{(K_0(a/\lambda)+K_2(a/\lambda))/2}$. Provided $a \sim 3
\lambda$ in the experiments, $l_{\rm FS}$ shortens with either
decreasing $t$ or $1/a$. In the studied samples with $t \sim
5-30$\,$\mu$m,~\cite{Fasano2003}  $l_{\rm FS}\approx t/10 \sim a$,
and then this dimensional crossover is quite likely at the origin of
the observation of a non-vanishing structure factor for large
wavelengths.  For thick enough samples and/or smaller $a$ such that
$l_{\rm FS} \gtrsim 10\,a$, the crossover to disordered hyperuniform
vortex density fluctuations might be observed experimentally. This
would be a state with directional
hyperuniformity.\cite{torquato2016}  In other words, vortex matter
nucleated in thin samples with a dense distribution of planar
defects effectively behave as a collection of one-dimensional
elastic manifolds. The suppression of hyperuniformity in elastic
structures nucleated in media with planar disorder as identified in
this work may be also observed in a broad spectrum of systems with a
control parameter tuning the dimensional crossover. In addition, a
direct mapping between the thickness of a classical system and the
temperature in quantum systems can be made, signposting the
conditions for disordered hyperuniformity to persist in
planarly-confined quantum systems such as optical traps.  These
results are rather important on the search for novel disordered
hyperuniform classic and quantum materials presenting exotic
physical properties.

\section{Methods}

\subsection*{Sample preparation and characterization}

We studied Bi$_2$Sr$_2$CaCu$_2$O$_{8+\delta}$ and
YBa$_{2}$Cu$_{3}$O$_{7}$ samples grown and characterized by means of
X-ray diffraction, transport and magnetometry techniques. The
Bi$_2$Sr$_2$CaCu$_2$O$_{8+\delta}$ samples with point disorder and
few planar defects were grown by means of the flux method and have a
$T_{\rm c} \sim 90$\,K; further details on the crystallographic and
superconducting properties of these samples can be found in
Ref.\,\onlinecite{Correa2001}. The
Bi$_2$Sr$_2$CaCu$_2$O$_{8+\delta}$ samples with many planar defects
were grown following the travelling-solvent-floating-zone method
using an image furnace with two ellipsoidal mirrors and has a
critical temperature of $\sim 87$\,K. The YBa$_{2}$Cu$_{3}$O$_{7}$
single crystals were obtained following a growth from the melt
technique and are fully oxygenated with $T_{\rm c} \sim 92$\,K, see
Ref.\,\onlinecite{Delacruz1994} for further details on the growing
method.

\subsection*{Vortex imaging by means of magnetic decoration experiments}

We image individual vortex positions at the sample surface in large
fields-of-view ranging from 1,000 to 35,000 vortices by means of
magnetic decoration  experiments.~\cite{Fasano2005} For all the data
presented here, the magnetic field is applied above $T_{\rm c}$ and
the sample is cooled down to 4.2\,K. At this temperature Fe
particles are evaporated in a pressure-controlled helium chamber and
clusters of these particles land on the sample surface decorating
the positions of vortices. Even though the snapshots of the
structure are taken at 4.2\,K, during the field-cooling process the
vortex structure gets frozen, at length-scales of the lattice
parameter $a_{\rm 0}$, at a temperature $T_{\rm freez} \sim T_{\rm
irr}$, the irreversibility temperature at which pinning (sample
disorder) sets in.~\cite{CejasBolecek2016} On  further cooling down
to 4.2\,K,  vortices can move but in length scales of the order of
coherence length, 200 times smaller than the typical size of a
vortex as detected by magnetic decoration. Therefore the structure
imaged by magnetic decoration at 4.2\,K  corresponds to the
equilibrium one at $T_{\rm freez}$.

\subsection*{Structure factors}

In order to calculate the structure factors we start considering the
vortex density modulation
\begin{equation}
\rho(x,y,z)=\frac{1}{t} \Sigma_{j=1}^{N_{\rm v}} \delta(x - x_{
j}(z))\delta(y - y_{j}(z))-\rho_0.
\end{equation}
where $\rho_0$ is the average density and $N_{\rm v}$ the number of
vortices. In magnetic decoration experiments we have access to the
vortex structure at the surface, namely $\rho(x,y,z=0)$. The
structure factor is obtained from the two-dimensional Fourier
transform of the density, $\hat{\rho}$, as
\begin{equation}
S({\bf q})= |\hat\rho(q_x,q_y,z=0)|^2.
\end{equation}
In the same token, the one-dimensional structure factor in a vortex
row, $S_{1}$ is obtained from the vortex density modulation along a
line $\hat{\rho_{1}}$. The angular structure factor $S_{\theta}(q)$
is the polar-coordinate representation of $S(q_x,q_y)$, see
Fig.\,\ref{fig:Figure2} (f) for schematics.

The angularly-averaged $\langle S(q) \rangle$ has to be calculated
carefully when studying the  low-$q$ density modes. Due to finite
size effects, the borders and shape of the experimental
field-of-view hinder the study of $S(\mathbf{q})$ in the low-$q$
range due to the annoying windowing effect. In rectangular
fields-of-view as we study here, this artifact produces an excess in
$S(q_{\rm x}, q_{\rm y} )$ in a cross-shaped region centered at
$q_{\rm x}=q_{\rm y}=0$.  When analyzing our experimental data, in
order to get rid of this effect we neglect the contribution from
this cross.  In simulations, this effect is avoided considering
in-plane periodic boundary conditions.

\subsection*{Computer simulation details}

The numerical simulations performed here consider an overdamped
Langevin dynamics at a temperature $T$
\begin{eqnarray}
&\eta& \partial_\tau {\bf r}_{i}(z) = -\frac{\delta E}{\delta {\bf r}_{i}(z)} + \xi({\bf r}_{i}(z),\tau) \\
&\langle& \xi({\bf r}_{i},\tau) \xi({\bf r}_{j}',\tau') \rangle = 2
\eta k_{\rm B} T \delta_{ij}  \delta(\tau-\tau')
\end{eqnarray}
where $\tau$ is the time and $\eta$ is the Bardeen-Stephen friction.
In order to emulate the experimental conditions we have simulated
systems with densities such that $a =3 \lambda_{\rm ab}$. The values
of the pinning strength of planar defects, $A_{\rm pin}$, and the
number of planar defects, $N_{\rm d}$, were tuned such that the
vortex system displays at the surface of the simulated sample a
structure qualitatively similar to the one observed in magnetic
decoration experiments. We have used $L_{\rm z}=64$, $N_{\rm v}=40$
and $a_0=3\lambda_{\rm ab}$, $A_{\rm pin}=0.2\epsilon_0$, $k =
\epsilon_0/\lambda_{\rm ab}^2$ , $N_{\rm d}=38$,
$L_{\parallel}=60\lambda_{\rm ab}$, $L_{\perp}=L_{\parallel}
\sqrt{3}/2$. In order to mimic the experiments we start the
simulation at $T=0.5 \epsilon_0 \lambda_{\rm ab}/k_{\rm B}$, reduce
$T$ slowly, and equilibrate the system at $T=0.001 \epsilon_0/k_{\rm
B}$. When the averaged structure factor appears to be stationary, we
analyze different properties of the final configuration and average
them over several realizations of the same protocol.

\subsection*{Planar elastic vortex array model: Analytical details}

The simple model of a planar elastic vortex array considered in the
discussion is described in the continuum by the scalar longitudinal
displacement field $u_{1}(s_{\parallel},z)$ with hamiltonian
\begin{equation}
{\cal H} \approx \int {dq_{\rm z} dq_{\parallel}}
|\hat{u}_{1}(q_{\parallel},q_{\rm z})|^2 [c_{11}(q_{\parallel},
q_{\rm z})q_x^2+c_{44}(q_{\parallel}, q_{\rm z})q_{\rm z}^2],
\label{eq:hamiltonian}
\end{equation}
where $\hat{u}_{1}(q_{\parallel},q_{\rm z})$ is the Fourier
transform of $u_{1}(s_{\parallel},z)$ and
$c_{11}(q_{\parallel},q_{\rm z})$ and $c_{44}(q_{\parallel},q_{\rm
z})$ are the dispersive compression and tilt elastic modulii. At
thermal equilibrium
\begin{equation}
\langle |\hat{u}_{1}(q_{\parallel},q_{\rm z})|^2 \rangle =
\frac{k_{\rm B} T}{ c_{11}(q_{\parallel}, q_{\rm
z})q_{\parallel}^2+c_{44}(q_{\parallel}, q_{\rm z})q_{\rm z}^2 }.
\end{equation}
The modulation of the coarse-grained vortex density at
long-wavelengths is
\begin{equation}
\delta n(s_{\parallel},z) \approx -n_0 \partial_{s_{\parallel}}
u_{1}(s_{\parallel}, z),
\end{equation}
where $n_0$ is the average number of vortices per unit length along
the vortex row of the planar vortex array. Then, for small
$q_{\parallel}$,
\begin{equation}
n_0 {S_1}(q_{\parallel}, q_{\rm z}) \equiv \langle |\delta
\hat{n}(q_{\parallel},q_{\rm z})|^2 \rangle \approx n_0
q_{\parallel}^2 \langle |\hat{u}_{1}(q_{\parallel},q_{\rm z})|^2
\rangle.
\end{equation}
By assuming translational invariance along $z$ and evaluating the
elastic constants at $q_{z} = 2\pi/t$, the correlation function
\begin{eqnarray}
    {S_1}(q_{\parallel}, z_1-z_2)
    &=& n_0^{-1}\langle \delta \hat{n}(q_{\parallel},z_1) \delta \hat{n}^*(q_{\parallel},z_2) \rangle \nonumber \\
    &\approx& \frac{n_0 k_{\rm B} T e^{-|z_1-z_2|/\xi_{\parallel}(q_{\parallel})}}{c_{11}(q_{\parallel},2\pi/t) \xi_{\parallel}(q_{\parallel})} ,
\end{eqnarray}
with
\begin{equation}
\xi_{\parallel}(q_{\parallel})=q_{\parallel}^{-1}\sqrt{c_{44}(q_{\parallel},2\pi/t)/c_{11}(q_{\parallel},2\pi/t)}
\end{equation}
the correlation length along the $z$-direction. Neglecting surface
effects,~\cite{Marchetti1993} the structure factor at the sample
surface is $S_1(q_{\parallel})\equiv S_1(q_{\parallel},z_1-z_2=0)$.
Then, for $\xi_{\parallel}(q_{\parallel})<t$ we get
Eq.\;(\ref{eq:1lane}). Finite-size effects appear when
$\xi_{\parallel}(2\pi/l_{\rm FS})=t$, obtaining the crossover length
$l_{FS}$ of Eq.(\ref{eq:yfs}).

When weak point disorder is added, the dimensional crossover still
exists and the effective rigid-vortex chain regime for
$q_{\parallel}<2\pi/l_{\rm FS}$ is equivalent to an elastic
interface in a random-periodic type of disorder. In this case, at
equilibrium~\cite{bustingorry2010}
\begin{eqnarray}
\langle{|\hat{u}_1(q_{\parallel})|^2}\rangle &\sim&
q_{\parallel}^{-(1+2\zeta_{\parallel})}
\end{eqnarray}
with $\zeta_{\parallel}=1/2$. Since $S_1(q_{\parallel}) =
q^2_{\parallel} \langle{|\hat{u}_1(q_{\parallel})|^2}\rangle$, the
second regime of Eqs. (\ref{eq:1lane}) and (\ref{eq:1lanew}) is
obtained for $q_{\parallel} < 2\pi/l_{\rm FS}$ or $s_{\parallel} >
l_{\rm FS}$. The correlations in the infinite-thickness limit can be
tackled analytically by mapping to the Cardy-Ostlund
model.~\cite{nattermann2000} In this case $W$ displays subtle
additive $\log^2 (s_{\parallel})$ corrections to the
$\log(s_{\parallel})$ growth,~ \cite{nattermann2000} implying
\begin{eqnarray}
S_1(q_{\parallel} \to 0)\sim -q_{\parallel}\log(q_{\parallel}) \to
0.
\end{eqnarray}
Then, in this limit the system is class III hyperuniform in contrast
with the class II hyperuniformity found in the clean case.

Once the corresponding formulas for $S_1(q_{\parallel})$ are known,
the displacement correlator of Eq.\,(\ref{eq:1lanew}) of the main
text can be obtained considering that

\begin{eqnarray}
W &=& \langle [u_{1}(s_{\parallel}, z)-u_{1}(0, z)]^2
\rangle-\langle [u_{1}(s_{\parallel}, z)-u_{1}(0, z)] \rangle^2
\nonumber \\
&\sim& \int {d q_{\parallel}}\;\; q_{\parallel}^{-2}
S_1(q_{\parallel})(1-\cos(q_{\parallel}s_{\parallel})).
\end{eqnarray}

\section*{Data availability}
All relevant data are available from the authors upon request.

\section*{Code availability}
All relevant code for simulations are available from the authors
upon request.

\section*{References}

\bibliography{biblio}

\section*{Acknowledgements}
We thank Thierry Giamarchi for stimulating discussions.

This work was supported by the Argentinean National Science
Foundation (ANPCyT) under Grants PICT 2017-2182 and PICT 2018-1533;
by the Universidad Nacional de Cuyo research grants 06/C566 and
06/C575; and by Graduate Research fellowships from CONICET for J. R.
P., F. E.,  J. A. S., R. C. M., G. R. and N. R. C. B.

\section*{Author contributions}

Y.F. and A.B.K. designed research and discussed the general method
to analyze the data, Y. F. and R.C.M. performed measurements, G.N.
and P. P. grew samples, F. E. and A. B. K. performed simulations and
theoretical calculations, J. R. P., F. E., J.A.S., G. R.,  A. B. K.
and Y.F. analyzed data; all authors discussed the data analysis and
interpretation; Y.F. and  A.B.K. wrote the paper.

\section*{Competing interests}
The authors declare no competing interests.

\section*{Additional information}

\textbf{Correspondence} and requests for materials should be
addressed to Y.F.

\end{document}